\documentclass[aps,prl,twocolumn,showpacs,superscriptaddress]{revtex4-1}
\usepackage{graphicx}
\usepackage{amssymb}
\usepackage{epstopdf}
\begin{document}
\preprint{}

\title{Probing the Sensitivity of Electron Wave Interference \\ to Disorder-Induced Scattering in Solid-State Devices}

\author{B.~C.~Scannell}
\affiliation{Department of Physics, University of Oregon, Eugene, OR 97403-1274, USA}

\author{I.~Pilgrim}
\affiliation{Department of Physics, University of Oregon, Eugene, OR 97403-1274, USA}

\author{A.~M.~See}
\affiliation{School of Physics, University of New South Wales, Sydney, NSW, 2052, Australia}

\author{R.~D.~Montgomery}
\affiliation{Department of Physics, University of Oregon, Eugene, OR 97403-1274, USA}

\author{P.~K.~Morse}
\affiliation{Department of Physics, University of Oregon, Eugene, OR 97403-1274, USA}

\author{M.~S.~Fairbanks}
\affiliation{Department of Physics, University of Oregon, Eugene, OR 97403-1274, USA}

\author{C.~A.~Marlow}
\affiliation{Department of Physics, University of Oregon, Eugene, OR 97403-1274, USA}

\author{H.~Linke}
\affiliation{Division of Solid State Physics and Nanometer Structure Consortium (nmC@LU), Lund University, Box 118, S-221 00, Lund, Sweden}

\author{I.~Farrer}
\affiliation{Department of Physics, Cavendish Laboratory, J. J. Thompson Avenue, Cambridge, CB3 0HE, United Kingdom}

\author{D.~A.~Ritchie}
\affiliation{Department of Physics, Cavendish Laboratory, J. J. Thompson Avenue, Cambridge, CB3 0HE, United Kingdom}

\author{A.~R.~Hamilton}
\affiliation{School of Physics, University of New South Wales, Sydney, NSW, 2052, Australia}

\author{A.~P.~Micolich}
\affiliation{School of Physics, University of New South Wales, Sydney, NSW, 2052, Australia}

\author{L.~Eaves}
\affiliation{School of Physics and Astronomy, University of Nottingham, Nottingham NG7 2RD, United Kingdom}

\author{R.~P.~Taylor}
\affiliation{Department of Physics, University of Oregon, Eugene, OR 97403-1274, USA}

\date{\today}

\begin{abstract}
The study of electron motion in semiconductor billiards has elucidated our understanding of quantum interference and quantum chaos. The central assumption is that ionized donors generate only minor perturbations to the electron trajectories, which are determined by scattering from billiard walls. We use magnetoconductance fluctuations as a probe of the quantum interference and show that these fluctuations change radically when the scattering landscape is modified by thermally-induced charge displacement between donor sites. Our results challenge the accepted understanding of quantum interference effects in nanostructures.
\end{abstract}


\maketitle

\section{I. Introduction}

The use of solid-state devices to replicate electron interference effects originally observed in vacuum holds considerable interest for fundamental and applied physics. The Aharonov-Bohm (AB) effect serves as an example \cite{AharonovPR59,ChambersPRL60}. If a ring-shaped device enclosing an area $A$ is connected on opposing sides to source and drain leads, then an electron flowing into the ring can take the clockwise or anticlockwise arm to exit at the other side. At temperatures low enough to establish electron phase coherence, quantum interference between these paths leads to a sinusoidal oscillation in conductance as a function of magnetic field $B$ applied perpendicular to the ring with a period $\Delta B = h/eA.$ This geometry-induced AB effect was first observed using gold rings \cite{WebbPRL85}. Subsequent measurements of semiconductor wires \cite{KaplanPRL86,SkocpolPRL86,ThorntonPRB87,TaylorSurfSci88} focused on AB effects generated by material-induced scattering events. In these diffusive devices, a multitude of AB `loops' arise from the many possible paths an electron can follow through the device due to scattering both from the device walls and from electrostatic perturbations induced by ionized donors. The resulting magnetoconductance fluctuations (MCF) have a spectral content \cite{TaylorSurfSci88} determined by the distribution of loop areas and can therefore be used as a magnetofingerprint of the electron dynamics \cite{LeePRL85}. Semi-classical theory developed for diffusive devices emphasizes the crucial role played by donor perturbations by predicting that the magnetofingerprint can be changed significantly by changing the location of just one scattering site \cite{FengPRL86,FengPRB93}. Experiments performed on diffusive wires have since confirmed the sensitivity of MCF to the movement of these scatterers \cite{GusevJPhysCondMat89,HeinzelPRB00,KlepperPRB91,TaylorICPS88,YinSSC93}. 

Ultrapure semiconductor heterostructures employ the modulation doping technique \cite{DingleAPL78} to spatially separate the donors from the two-dimensional electron gas (2DEG) that forms at the heterostructure interface. This reduces the magnitude of the donor perturbations induced in the 2DEG's electrostatic landscape. Whereas the strong electrostatic perturbations induced by `in-plane' donors in diffusive devices generate large-angle scattering events, these weaker perturbations are more likely to generate small-angle scattering. Because electron mobility is naturally weighted towards large angle-scattering \cite{JuraNatPhys07,ColeridgePRB91}, the associated mean free path $\ell_{\mu}$ can exceed the device size $L$, satisfying the traditional definition of a semiconductor `billiard' ($L < \ell_{\mu}$) \cite{BeenakkerBook91}. However, this definition does not exclude the possibility of small-angle scattering within the device. It is not therefore a good gauge of ballistic transport, which requires electrons to follow straight trajectories between specular reflections from the device walls. This is highlighted by recent scanning gate microscopy (SGM) studies \cite{JuraNatPhys07,TopinkaNat01,CrookPRL03} which show that small-angle scattering produces branching of trajectories at length scales much shorter than the mean free path for large-angle scattering. In particular, SGM performed on heterostructures with electron mobilities in excess of those used for fabricating billiards indicate that significant small-angle branching can be expected within semiconductor billiards \cite{JuraNatPhys07}. 

These SGM studies re-ignite a central question for quantum transport research---how does the presence of small-angle scattering impact the AB effect in semiconductor billiards? The semi-classical theory developed for the diffusive regime does not quantify the minimum perturbation size required for its applicability \cite{FengPRL86}. Consequently, it is not clear whether the sensitivity of the AB effect to perturbations highlighted by theoretical \cite{FengPRL86} and experimental \cite{GusevJPhysCondMat89,HeinzelPRB00,KlepperPRB91,TaylorICPS88,YinSSC93} investigations of diffusive devices extends to the smaller perturbations experienced by billiards. The pioneering experimental \cite{MarcusPRL92,BerryPRB94,TaylorPRL97,TaylorPRB97,SachrajdaPRL98} and theoretical \cite{JalabertPRL90} investigations of billiards assumed that donor perturbations did not significantly affect the electron dynamics established by the scattering imposed by the device walls. The experimental results presented here challenge this picture by demonstrating that donor-induced small-angle scattering is the dominant factor influencing the MCF.

We employ two techniques to relocate charge between donor sites---thermal activation and illumination with a LED. Thermal activation \cite{TaylorCJP92,TaylorICPS88,EchternachPRB91} and illumination \cite{KlepperPhysicaB90,KlepperPRB91,DavisonPhysicaB90} have been used previously to demonstrate that charge relocation induces changes in the MCF for diffusive devices dominated by large-angle scatterers. In addition, illumination is a well-established technique for studying charge activation energies for donors \cite{TheisIPCS87,MooneyJAP90}. Here, we apply these techniques to investigate the impact of charge relocation on the billiard's MCF. We show that redistribution of the ionized donors reconfigures the MCF despite the billiard's fabricated geometry remaining unchanged. This reappraisal highlights the challenges of realizing genuine quantum ballistic transport in semiconductor nanostructures. Furthermore, the results address fundamental aspects of quantum interference and its sensitivity to electron scattering \cite{FengPRL86}. The MCF are observed to gradually reconfigure as the number of altered scattering sites in the billiard increases, demonstrating the cumulative impact of changing disorder on quantum interference processes. 

Our results also have broader implications beyond semiconductor physics. The quantum behavior of semiconductor billiards can be used to investigate whether universal characteristics are evident across diverse systems. For example, in quantum chaos studies, MCF were used to investigate the change from stable to chaotic electron dynamics for billiards with differently shaped walls \cite{MarcusPRL92,BerryPRB94} and for shape transitions within a single billiard \cite{TaylorPRL97,TaylorPRB97}. These studies, which neglected donor scattering, are influential in the broader research community due to analogous phenomena in quantum wells \cite{WilkinsonNat96}; in microwave \cite{StockmannPRL90}, optical \cite{GmachlSci98,WilkinsonPRL01}, and acoustic \cite{SchaadtPRE03} cavities; and cold atoms in optical traps \cite{ZhangPRL04}.

 \begin{figure}[h!]
 \includegraphics[width=1\columnwidth]{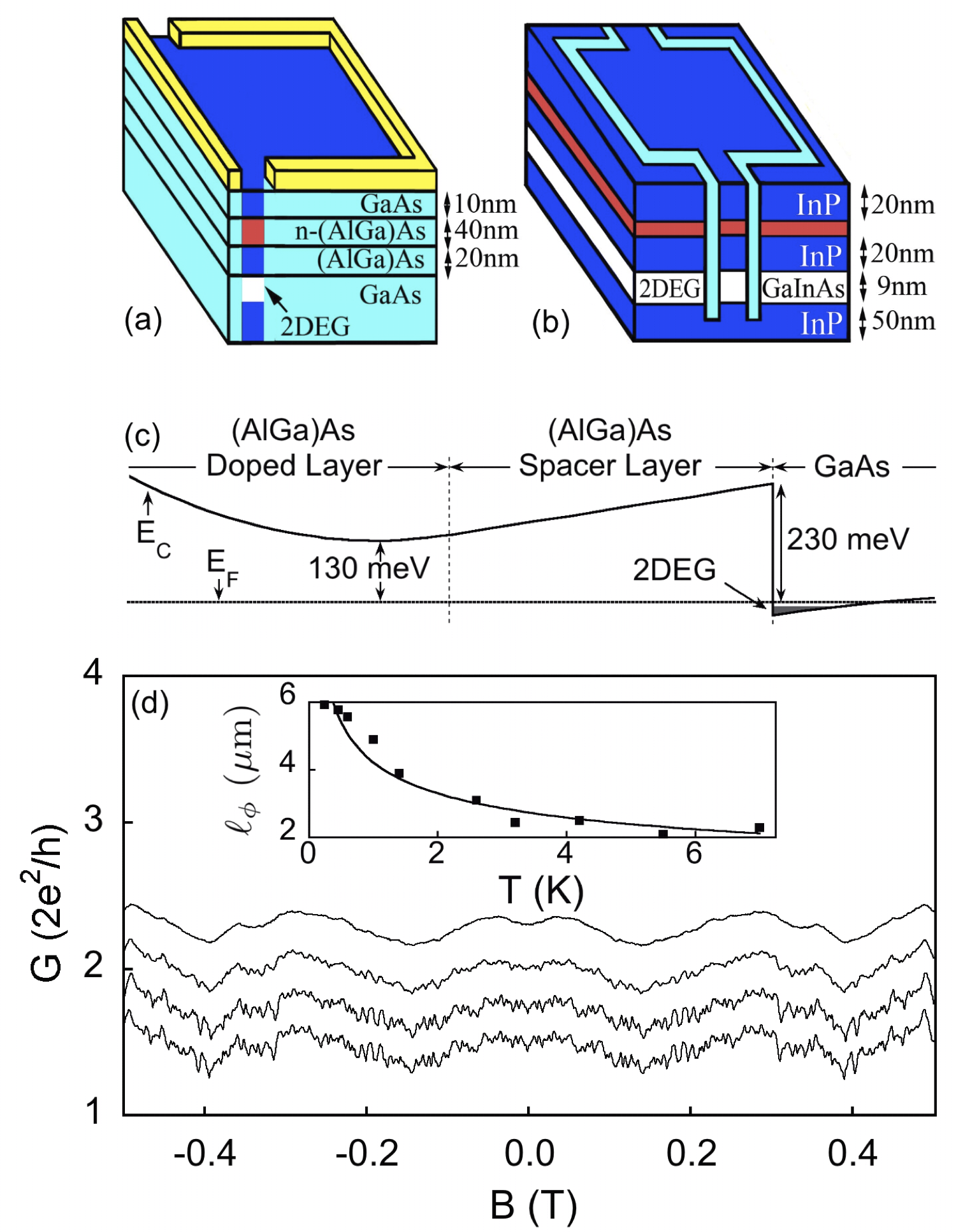}
 \caption{\label{f1}(Color online) Schematics of (a) GaAs/(AlGa)As and (b) (GaIn)As/InP billiards. The 2DEG (white) is located near the lower GaAs/(AlGa)As interface for (a) and in the (GaIn)As layer for (b). The modulation-doping (red) consists of a 40 nm (AlGa)As layer with a Si donor density of 1.2$\times$10$^{18}$ cm$^{-3}$  for (a) and a 1 nm InP layer with a Si density of 5$\times$10$^{18}$ cm$^{-3}$  for (b). The billiard walls (light blue) are defined by surface gates (yellow) in (a) and by etched trenches in (b). The heterostructure in (b) has a uniform top gate separated from the surface by a 1 $\mu$m insulating layer (not shown). (c) Simulation of the energy band structure for the GaAs/(AlGa)As heterostructure, plotting the conduction band edge $E_{C}$ v. distance from the 2DEG. (d) Magnetoconductance traces for a 1~$\mu$m (GaIn)As/InP billiard measured at temperatures $T$ of (top to bottom) 12 K, 5 K, 1 K, and 240 mK. The traces are offset for clarity. Inset: A plot of the electron phase coherence length $\ell_{\phi}$ as a function of $T$, and associated fit line, for the (GaIn)As/InP billiard.}
 \end{figure}

\section{II. Experiments}

To investigate the generic properties of electron interference in semiconductor billiards, we consider two widely studied material systems and two distinct methods for defining billiards. In each case, a 2DEG forms within a heterostructure (Fig. 1(a, b)). For the GaAs/Al$_{0.33}$Ga$_{0.67}$As system, electrostatic depletion generated by patterned surface gates defines `soft' billiard walls in the 2DEG (Fig. 1(a)) \cite{MicolichAPL02}. For the Ga$_{0.25}$In$_{0.75}$As/InP system, wet etching generates `harder' walls with steeper electrostatic gradients (Fig. 1(b)) \cite{MarlowPRB06}. The 2DEG densities are $n = 2\times10^{15}$ m$^{-2}$ and $7\times10^{15}$ m$^{-2}$, giving Fermi energies of $E_{F}$ = 8.4 meV and $50$ meV, respectively. Accordingly, the Fermi wavelengths (50 nm and 30 nm) are significantly smaller than the billiard widths of 1 $\mu$m, as required for semi-classical interference phenomena \cite{JalabertPRL90}. The fabrication of high-quality billiards involves the modulation doping technique shown in Fig. 1(c). Silicon donors are separated from the 2DEG plane by a spacer layer in order to minimize the electrostatic perturbations induced in the 2DEG by the ionized donors \cite{DingleAPL78,NixonPRB90}. Small-angle scattering induced by these perturbations is not sufficient to reduce $\ell_{\mu}$ to below the billiard width. The $\ell_{\mu}$ values for our billiards (3 $\mu$m and 6 $\mu$m for the GaAs/(AlGa)As and (GaIn)As/InP systems) are comparable to those measured in the pioneering MCF experiments \cite{MarcusPRL92,BerryPRB94}. 

Four-terminal conduction measurements were performed using AC lock-in techniques at a frequency of 37 Hz and an excitation current of 1 nA. The bottom trace of Fig. 1(d) shows the MCF generated by wave interference, superimposed on a classical conductance background for a square (GaIn)As/InP billiard measured at $T$ = 240 mK (base temperature). Billiards support a distribution of AB loop areas, leading to the observed range of frequencies in the magnetoconductance \cite{JalabertPRL90,MarcusPRL92,SachrajdaPRL98,MarlowPRB06}. Raising $T$ reduces the electron phase coherence length $\ell_{\phi}$ \cite{BirdPRB95,MarlowPRB06}: by $T$ = 12 K, $\ell_{\phi}$ is smaller than the billiard width, suppressing the MCF and leaving only the classical background.  

To investigate the impact of the donors on the AB loops, the billiards were cooled to base temperature in the dark and then warmed to temperatures high enough to supply the thermal energy necessary to relocate electron charge between dopant sites. Within the 1 $\mu$m$^2$ area of the (GaIn)As/InP and GaAs/(AlGa)As billiards, there are approximately 30,000 and 50,000 of these donor sites, respectively. Note, however, that these donors are distributed at various heights in the 40 nm modulation doped layer for the GaAs/(AlGa)As heterostructure. This introduces considerable spatial overlap of the donors such that the associated electrostatic perturbations defined within the 2DEG generate a much smaller number of individual scattering events sites within the billiard. Simulations of GaAs/(AlGa)As heterostructures similar to ours indicate that the electrostatic landscape comprises collective features on the scale of 0.1 $\mu$m, indicating approximately 100 scattering sites within the billiard \cite{NixonPRB90}. If the trajectories forming the AB loops are determined purely by the fabricated geometry of the billiard walls, as traditionally believed, then the MCF will be immune to this shift in charge between dopants. 

We annealed the billiard devices by warming in the dark to an intermediate temperature $T_{i}$ for a fixed time $t$~=~30 minutes and then cooled the billiards back to base temperature. We then compared the pair of MCF traces taken at base temperature before and after this thermal cycle. Figure 2 shows three pairs of MCF traces measured on the (GaIn)As/InP billiard. The bottom pair of traces were recorded back-to-back (i.e., $T_{i}$ = 240 mK) and demonstrate the well-known reproducibility of MCF traces at low temperatures. The middle pair of traces correspond to $T_{i}$  = 115 K and reveal the same high degree of correlation as the bottom pair. However, the top pair have visibly decorrelated after thermally cycling to $T_{i}$  = 300 K. Our equivalent MCF measurements performed on the GaAs/(AlGa)As billiard, shown in Fig. 3, demonstrate that this decorrelation is a generic effect for semiconductor billiards. We note that the MCF of the GaAs/(AlGa)As billiard have a smaller high-frequency component than the (GaIn)As/InP billiard due to the billiard's smaller $\ell_{\phi}$ of 3.6 $\mu$m at $T = 240$ mK \cite{MarlowPRB06}.

 \begin{figure*}[t]
 \includegraphics[width= 1.5\columnwidth]{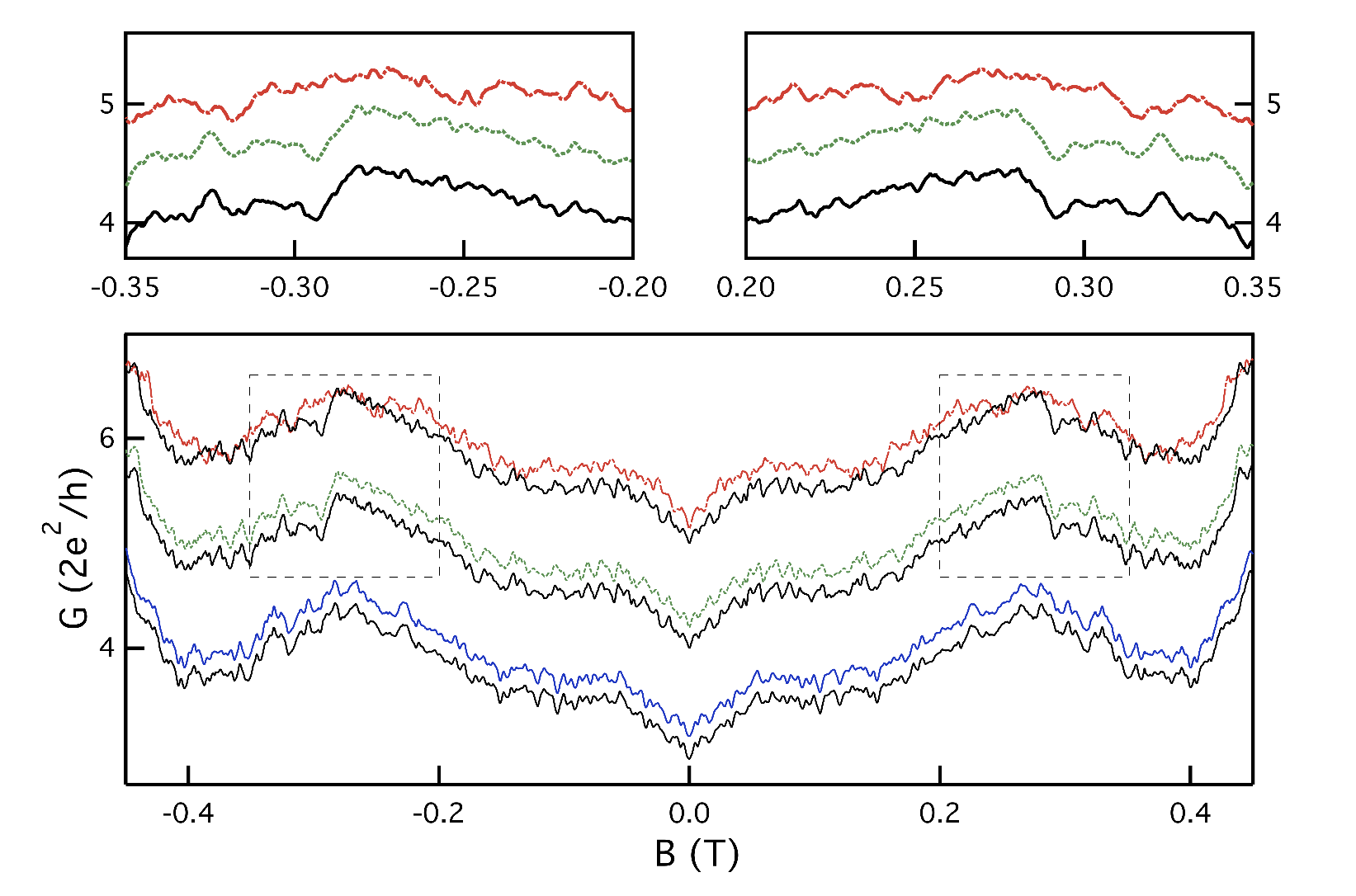}
 \caption{\label{f2}(Color online) Comparison of MCF for the (GaIn)As/InP billiard after being warmed to three intermediate temperatures $T_{i}$. Top to bottom, the $T_{i}$ values for the pairs of traces are 300 K, 115 K, and 240 mK. The traces are offset for clarity. Top: Magnified comparison of traces taken with $T_{i}$ = 115 K (lower [black] and middle [green]) and $T_{i}$ = 300 K (lower [black] and upper [red]). }
\end{figure*}

To quantify the change in MCF as a function of $T_{i}$, we applied the following correlation function \cite{TaylorICPS88,TaylorCJP92} to each pair of traces $G_{1}(B)$ and $G_{2}(B)$:

\begin{eqnarray}
F &=& \sqrt{ 1 - \frac{ \left\langle \left[ G_{1}(B) - G_{2}(B) \right]^{2} \right\rangle_{B}}{N} },~~~\text{where} \\
N &=& \left\langle \left[ G_{x}(B) - G_{y}(B) \right]^{2} \right\rangle_{B}.
\end{eqnarray}

The symbol $\langle~\rangle_{B}$ represents an average over the data points spanning $-B_{c} < B < B_{c}$, where $B_{c}$ is the field at which the cyclotron diameter matches the billiard width. For the GaAs/(AlGa)As billiard, $B_{c} = 0.16$ T, and 640 data points fall in this range; for the (GaIn)As/InP billiard, $B_{c} = 0.28$ T, and 1120 data points fall in this range. The normalization constant $N$ is calculated by averaging the correlations of 15 pairs of traces ($G_{x}(B)$ and $G_{y}(B)$) that have been thermally cycled to $T_{i}$ = 300 K. Adopting this normalization, the correlation scale varies between 1 for mathematically identical traces to 0 for decorrelation induced by a $T_{i}$ = 300 K cycle. We note that the form of the correlation function minimizes noise in $F$ for values close to 1. This is important for obtaining an accurate fit to the dependence of $F$ on $T_{i}$ (see below).  

In Fig. 4, we show the results for the (GaIn)As/InP and GaAs/(AlGa)As billiards. For comparison, we show previously reported decorrelations of the MCF for a GaAs wire \cite{TaylorCJP92,TaylorICPS88}. The wire was 10 $\mu$m long, 50 nm high, and 90 nm wide, and was heavily doped with silicon ($n = 5 \times 10^{24}$ m$^{-3}$, $E_{F}$ = 128 meV) distributed uniformly throughout the wire's cross-section. Due to the absence of modulation doping, the strong electrostatic perturbations of these `in-plane' dopants generates large-angle scattering, resulting in $\ell_{\mu}$ = 0.3 $\mu$m. The thermally-induced decorrelation of the MCF for this diffusive scattering system (strong perturbations) is clearly similar to that of the two modulation-doped billiards (weak perturbations).

 \begin{figure*}[t!]
 \includegraphics[width= 1.5\columnwidth]{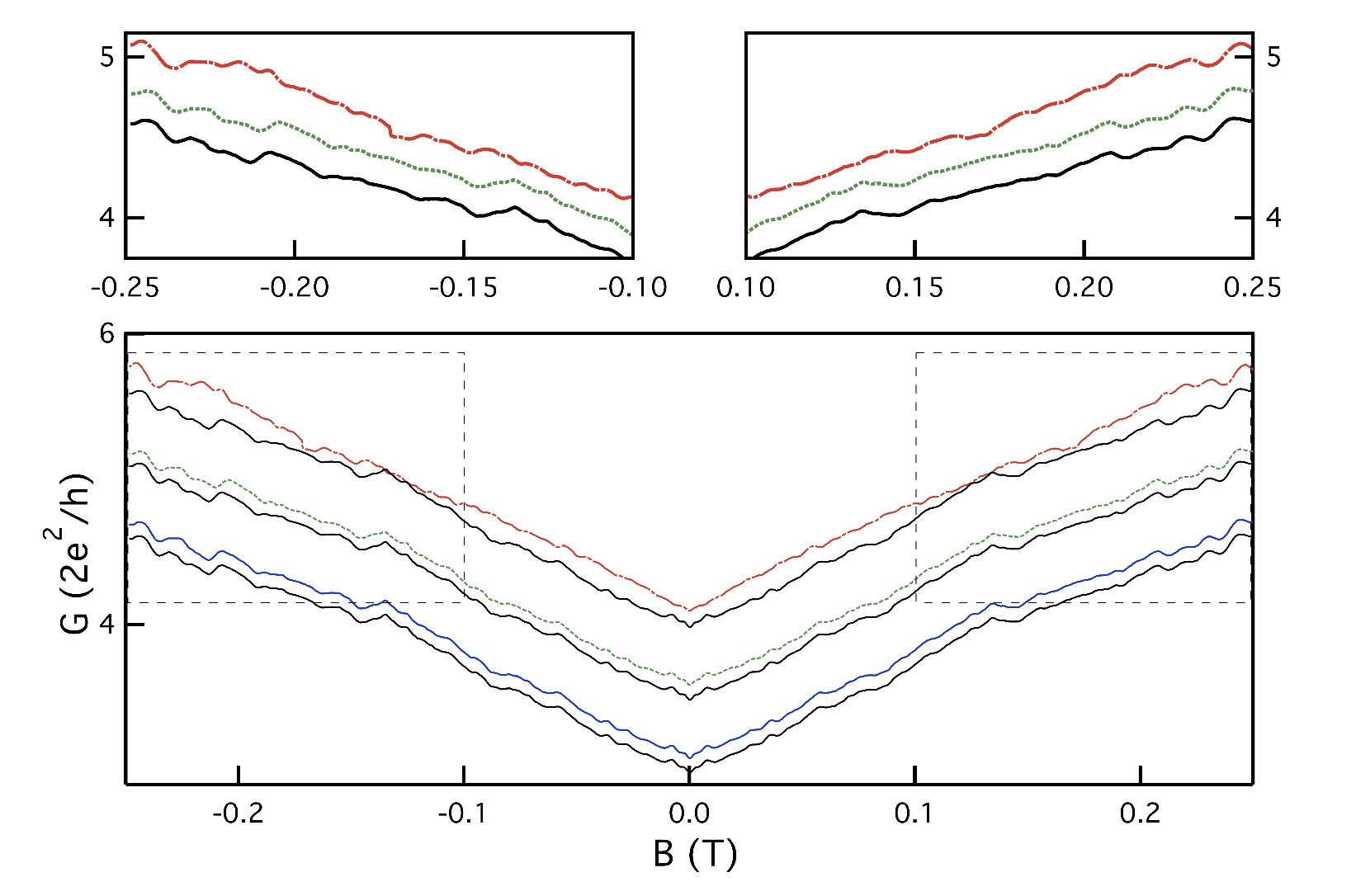}
 \caption{\label{f2}(Color online) Comparison of MCF for the GaAs/(AlGa)As billiard after being warmed to three intermediate temperatures $T_{i}$. Top to bottom, the $T_{i}$ values for the pairs of traces are 300 K, 115 K, and 240 mK. The traces are offset for clarity. Top: Magnified comparison of traces taken with $T_{i}$ = 115 K (lower [black] and middle [green]) and $T_{i}$ = 300 K (lower [black] and upper [red]). }
\end{figure*}

\section{III. Discussion}

For all three systems, we attribute the MCF decorrelation to the relocation of electron charge among silicon dopants. This charge relocation changes the electrostatic landscape experienced by the electrons \cite{NixonPRB90,BervenPRB94}, which in turn reshapes the distribution of AB loops. The probability $P$ for this process depends upon the available thermal energy according to $P \sim \exp(- \beta / k_{B} T_{i})$, where $\beta$ represents the activation energy for charge relocation. The charge redistribution is also governed by the time $t$ the billiard is held at $T_{i}$. We model the decorrelation behavior using $F = \exp[-\eta \exp (- \beta / k_{B} T_{i})]$, where $\eta = t/\tau_{0}$ and $\tau_{0}$ is a characteristic time for charge transfer \cite{TaylorCJP92}. The decorrelation data are fitted with this formula in Fig. 4.  The insets in Fig. 4 are simulations demonstrating how the form of the fall-off depends upon $\beta$ and $\eta$: $\beta$ (upper) and $\eta$ (lower) are varied while keeping the other parameter fixed. The GaAs/(AlGa)As billiard's decorrelation commences at a higher $T_{i}$ than for the other systems, indicating a higher $\beta$ value. 

To explain this observation, consider the GaAs/(AlGa)As conduction band diagram of Fig. 1(c) in more detail. Electrons undergo thermally-assisted tunneling from the 2DEG through the energy barrier at the GaAs/(AlGa)As interface to the doped layer's conduction band edge. In this process, a thermal energy of ~130 meV is required for the electrons to reach barrier heights at which the barrier width is sufficiently narrow for tunneling. We note that the 20 nm tunneling distances required are similar to those observed in other tunneling structures \cite{NakagawaAPL86}. Having tunneled into the (AlGa)As conduction band, DX centers \cite{MaudePRL87} associated with the silicon dopants can then capture pairs of these electrons and change their charge state \cite{ChadiPRB89}. This DX capture energy is $\sim$250 meV for Al$_{0.33}$Ga$_{0.67}$As \cite{MooneyJAP90}. This sequential process provides a possible explanation for the high $\beta$ value of 350 $\pm$ 100 meV obtained from the fit.  In contrast, in the (GaIn)As/InP heterostructure, the doped layer's conduction band edge resides only 75 meV above the 2DEG \cite{MartinPhysE08}. Furthermore, the Si dopants are expected to form shallow traps 5.6 meV below the band edge \cite{Ramvall96}.  These smaller activation energies are consistent with the lower $\beta$ value observed for this system (45 $\pm$ 20 meV). Finally, the $\beta$ value of 80 $\pm$ 20 meV observed for the GaAs wire is close to the expected activation energy of 100 meV, which is set by the height of the DX center above $E_{F}$ for this layer \cite{TheisIPCS87}. We plan future MCF experiments to study the relocation of charge between donors and its detailed dependence on the entire thermal annealing cycle. The current experiments do not take into account the finite time to warm to $T_{i}$. Future investigations will take into account activation of charge not just at $T_{i}$ but also at the temperatures experienced during the warm-up and cool-down, which would allow more accurate $\eta$ values to be measured. 

\begin{figure}[t!]
\includegraphics[width=1 \columnwidth]{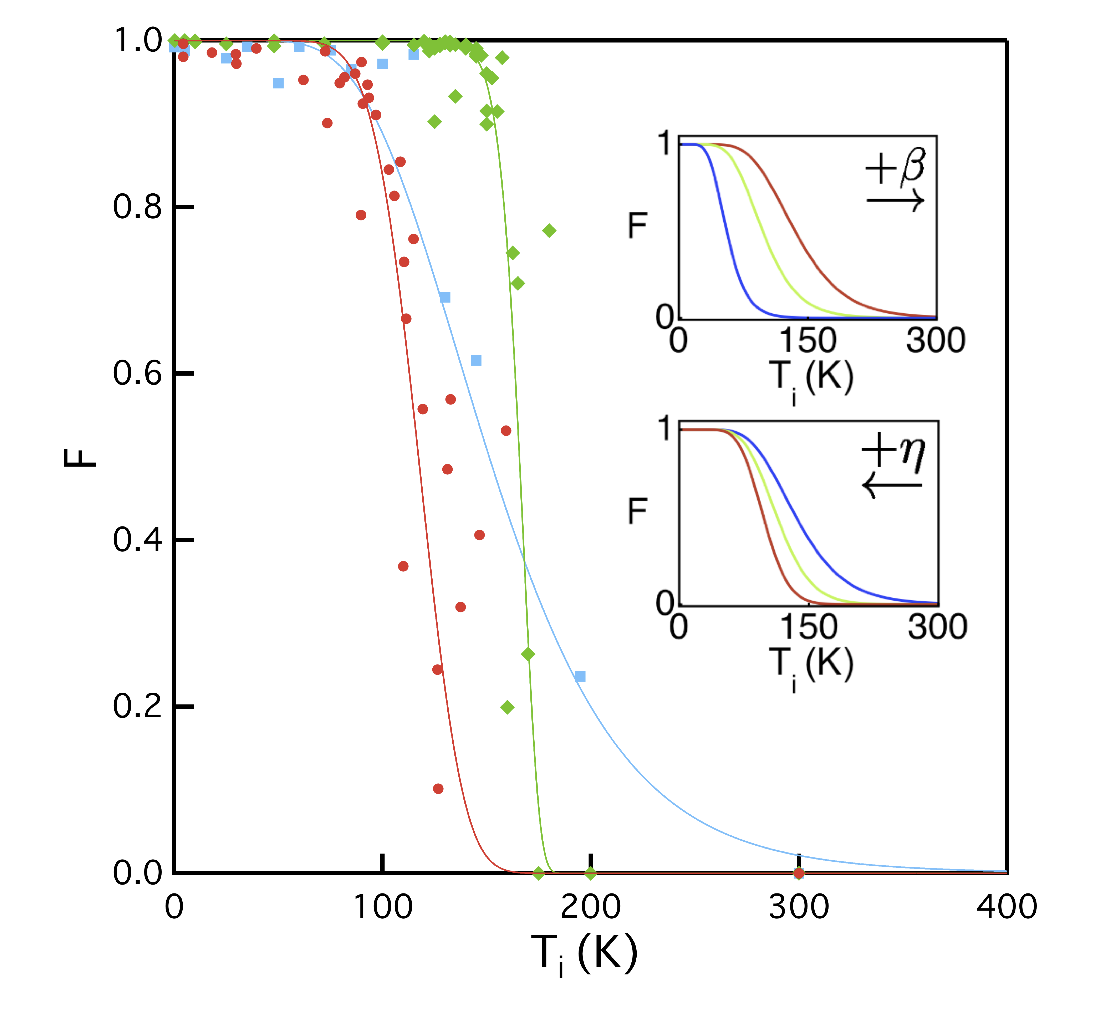}
\caption{\label{f4}(Color online) A plot of $F$ v. $T_{i}$ for the GaAs wire (red circles), the (GaIn)As/InP billiard (light blue squares), and the GaAs/(AlGa)As billiard (green diamonds). The insets show the annealing function's form as $\beta$ and $\eta$ are varied (see text).}
\end{figure}

\begin{figure}[t!]
\includegraphics[width=1.1 \columnwidth]{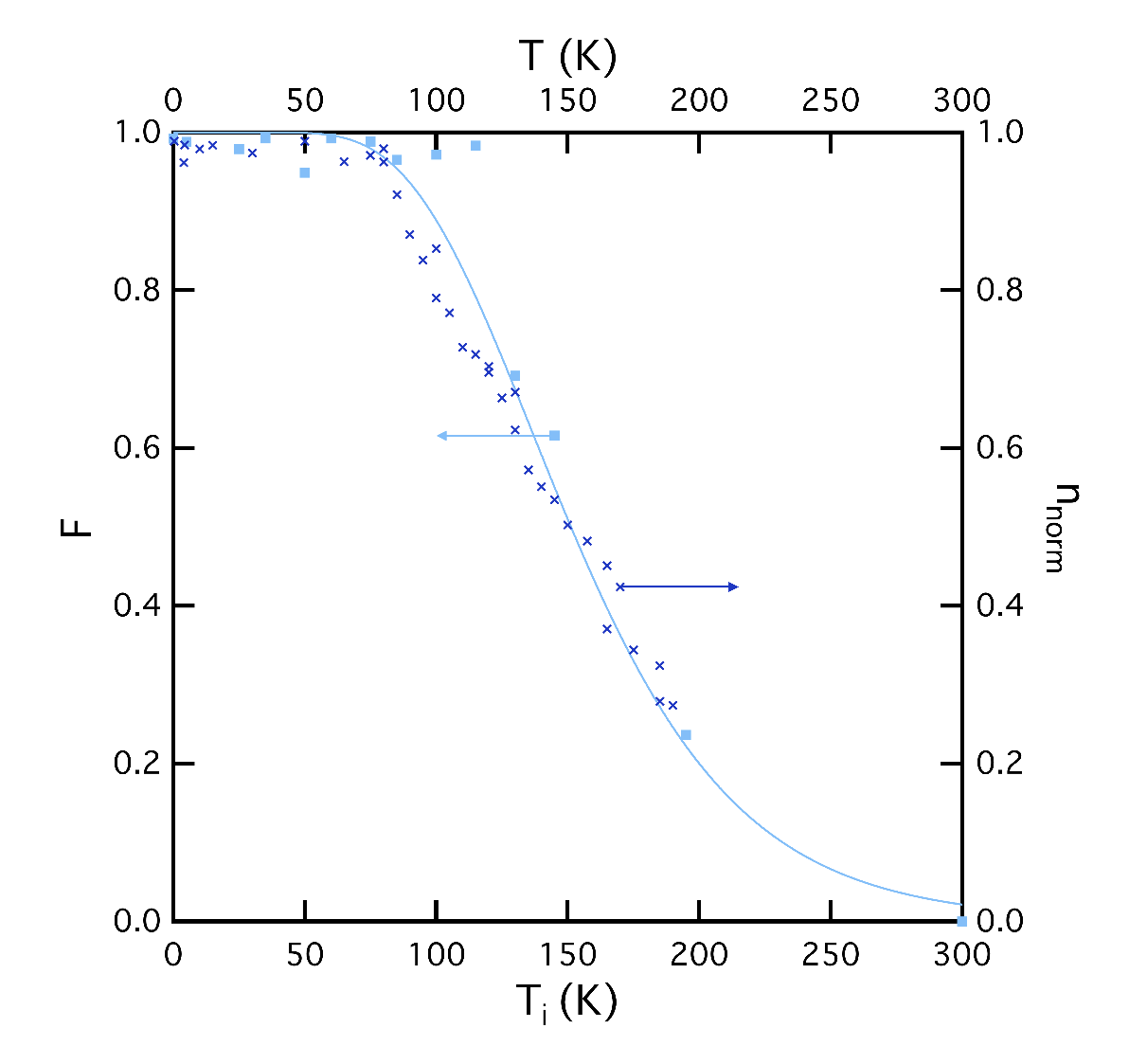}
\caption{\label{f4}(Color online) A plot of $F$ (left axis) v. $T_{i}$ (bottom axis) for the (GaIn)As/InP billiard (light blue squares), as well as $n_{\text{norm}}$ (right axis) v. $T$ (top axis) for the (GaIn)As/InP billiard (blue crosses). The decorrelation data for the (GaIn)As/InP are identical to those shown in Fig. 4.}
\end{figure}

To confirm the dependence of MCF decorrelation on the transfer of electrons between the 2DEG and donor sites, we employed a LED as follows \cite{TheisIPCS87}. After cooling to base temperature in the dark, the (GaIn)As/InP billiard was illuminated using a series of LED pulses. The resulting transfer of electrons from the dopants to the 2DEG was monitored with Hall measurements, which charted the rise in 2DEG density $n$ from the initial value of $n_{\text{min}}$ to the saturation value of $n_{\text{max}}$. The normalized $n$ value, $n_{\text{norm}} = (n - n_{\text{min}})/(n_{\text{max}}-n_{\text{min}})$, was then measured in the dark as a function of rising temperature. The decrease in $n_{\text{norm}}$ shown in Fig. 5 results from the thermally-activated transfer of electrons from the 2DEG back to the donors. The form of $n_{\text{norm}}$ v. $T$ is strikingly similar to that of $F$ v. $T_{i}$, highlighting the dependence of MCF decorrelation on charge transfer from the 2DEG.

We note that whether electrons have to tunnel from the 2DEG to the (AlGa)As layer in order to be captured by the donor sites, as observed here, depends on the precise layer architecture of the heterostructure.  For example, for heterostructures with much wider (AlGa)As layers than those employed for billiards research, parallel conduction can occur in the (AlGa)As layer \cite{MooneyJAP87}. Consequently, for these heterostructures, electron transfer across the interface barrier is not necessary to populate the (AlGa)As layer and the donors \cite{MooneyJAP87}. 

\section{IV. Conclusions}

In conclusion, we have shown that the electrostatic profile of ionized donors strongly affects the interference of electron waves in semiconductor billiards. By demonstrating this effect for two material systems and different fabrication methods, we propose this is a generic limitation of semiconductor billiards. This contradicts previous models in which the quantum dynamics was interpreted as being determined purely by the billiard walls \cite{MarcusPRL92,BerryPRB94,TaylorPRL97,TaylorPRB97,JalabertPRL90,LofgrenPRL04,SachrajdaPRL98} and which predict that the MCF should be robust to thermal cycling. In contrast to this traditional billiard picture, our experiments show that the modulation doping technique does not reduce the donor electrostatic perturbations sufficiently to ignore their presence. In particular, we have demonstrated experimentally that the sensitivity of the AB effect, highlighted by semi-classical theory developed for diffusive regime \cite{FengPRL86}, extends to billiards even though the size of the electrostatic perturbations in billiards is reduced significantly compared to diffusive systems. 

In addition to revealing the role of donor-induced disorder, our results also highlight central aspects of the sensitivity of quantum interference to electron scattering. The semi-classical theory predicts that changing one scattering site can have a significant effect on the MCF \cite{FengPRL86}, raising the question of whether one site change in the billiard is sufficient to completely decorrelate the MCF. The observed gradual decorrelation indicates that the total number of altered sites is important and has a cumulative effect on the MCF.

Our experiments can be combined with earlier observations demonstrating the lack of MCF sensitivity to billiard wall properties such as geometry and the softness of the wall's electrostatic profile \cite{MarlowPRB06,MicolichPRL01}. Taken together, these results can be seen as supporting an earlier proposal that the electron dynamics in billiards are generated by donor scattering \cite{MarlowPRB06}. Within this picture, the walls serve to repeatedly reflect electron trajectories towards the donor scattering sites \cite{MarlowPRB06}. However, we note that some MCF features, such as those induced by symmetries in the billiard walls \cite{LofgrenPRL04}, nevertheless display geometry dependences. Our results therefore highlight the need for quantitative investigations aimed at determining the relative contributions of walls and ionized donors to the MCF, which might be influenced by trajectory length and stability \cite{ReimannPRB97}, as well as dynamical effects such as quantum scarring \cite{AkisPRL97}.

In contrast to the sensitivity of the MCF to donor distributions, the classical magnetoconductance is expected to be determined predominantly by the billiard geometry. This difference in conduction processes originates from the sensitivity of the AB effect. The magnetic field dependence of the classical conduction is due to `focusing' of the trajectories into the billiard's entrance and exit due to curvature from the Lorentz force \cite{BeenakkerBook91}. Classical focusing is dependent only upon the opening by which the trajectories leave the billiard and is therefore relatively insensitive to small-angle scattering events. The AB effect is instead sensitive to the precise routes taken by the electrons through the billiard. Thus, whereas the presence of donors in the billiard will simply perturb the classical condition by broadening the focusing features, the donors will change the scattering loops substantially. This insensitivity to donors explains the previous success in observing classical ballistic effects \cite{BeenakkerBook91,FordPRL89,SpectorAPL90}.

Our observations demonstrating the importance of donors for quantum transport through billiards is different from earlier results performed on quantum point contacts, where the sensitivity to donors can be explained in terms of one-dimensional channels resulting from quantum confinement \cite{TimpPRB90,TaylorPRB92,HesslingJPhys95,KurdakPRB97}. These results on QPCs, along with our results on billiards, emphasize the central role that disorder plays for quantum transport in solid-state environments: the associated lack of reproducibility between devices and thermal cycles has implications not only for fundamental research but also for future nanoscale electronic devices. We expect this to apply to novel systems such as graphene \cite{GeimNatMater07,ChenNatPhys08} and carbon nanotubes \cite{CharlierRMP07}.  

\begin{acknowledgments}
Acknowledgments: D. K. Maude for useful discussions; the Office of Naval Research [N00014-07-0457], US Air Force [FA8650-05-1-5041], Australian Research Council [DP0772946, FT0990285, LX0882222], and Research Corporation for Science Advancement for funding; the Australian National Fabrication Facility; and nmC@LU.
\end{acknowledgments}

\end{document}